\documentclass[preprint,12pt]{elsarticle}

\usepackage{amssymb}
\usepackage{amsmath}
\usepackage{color}
\usepackage{amsthm}

\newcommand{\myvec}[1]{\vec{\mathbf{#1}}}

\journal{Journal of Magnetism and Magnetic Materials}

\begin{document}

\begin{frontmatter}



\title{Nonlinear dynamics in spin wave active ring oscillator (SWARO) driven near a dipole gap} 

\author[1]{Anirban Mukhopadhyay}
\affiliation[1]{organization={Department of Electrical Engineering, Indian Institute of Technology Madras},
            city={Chennai},
            postcode={60036}, 
            state={Tamil Nadu},
            country={India}}
            
\author[2,3]{Ihor I. Syvorotka}
\affiliation[2]{organization={Department of Crystal Physics and Technology, Scientific Research Company ``Electron-Carat"},
            city={Lviv},
            postcode={79031},
            country={Ukraine}}
\affiliation[3]{organization={Department of Semiconductor Electronics, Lviv Polytechnic National University},
            city={Lviv},
            postcode={79013},
            country={Ukraine}}
            
\author[1]{Anil Prabhakar}


\begin{abstract}
We investigate the nonlinear dynamics of spin wave active ring oscillators (SWAROs) injected with GHz drive signal.
The injected signal frequency was swept over a 5~MHz wide frequency range across a magnetostatic surface spin wave (MSSW) dipole gap, with the drive power varying from -10 to 10~dBm. 
We measured the output power spectra from the SWARO at different gains for each drive power and frequency combination. 
At higher drive amplitudes, the spin wave nonlinearity in the ring oscillator is suppressed, and the SWARO spectrum is pulled toward the drive frequency. 
Furthermore, near the drive frequency, we observe the formation of sidebands, which are the products of the nonlinear scattering processes. 
\end{abstract}


\begin{highlights}
\item We systematically study the nonlinear dynamics in spin wave active ring oscillators (SWAROs) when driven near a dipole gap.
\item The RF drive signal excites sidebands, observed in the output spectrum. We classify these sidebands into two types: type-A keeps a constant distance of 0.6~MHz from the drive signal, whereas type-B moves away from the drive frequency as it increases.
\item We achieve a spectral suppression of approximately 30~dB by increasing the drive power from -10 to 0~dB.
Additionally, large drive powers result in injection pulling.
\end{highlights}

\end{frontmatter}

\section{Introduction}
\label{sec:intro}
Spin waves (SWs) in ferrimagnetic films demonstrate interesting nonlinear behaviour such as modulation instability, periodic doubling, solitons, and chaos~\cite{prabhakar2009spin, Gurevich2020}.
Highly nonlinear dynamics can get excited in the neighbourhood of spin wave dipole gaps.
These gaps originate from the higher-order SW dispersion branches interacting with the lower ones~\cite{Schilz1972}.
In the gap, the higher branches have a positive dispersion coefficient ($D$), while the lower branches have a negative $D$~\cite{Wu2004gen}.
Hence, attractive and repulsive non-linear conditions can be achieved by shifting the operating frequency from one side to the other of the dipole gap.
Wu~\emph{et al.} excited self-modulational instability with both nonlinearities in an MSSW configuration near a dipole gap~\cite{Wu2004gen}.
Furthermore, coupled modulation instability was observed as interactions of two forward volume SWs, both excited near a dipole gap but with different frequencies and group velocities~\cite{wu2008}.
In both these studies, the instability produced solitons.
Furthermore, Prabhakar~\emph{et al.} studied the auto-oscillation phenomena near a dipole gap in a similar spin wave configuration~\cite{prabhakar1998auto}.

Similar nonlinear dynamics were studied in a spin wave active ring oscillator(SWARO)~\cite{WU2010163}.
A SWARO consists of a magnetic film acting as a frequency-selective nonlinear element and a variable gain amplifier.
The oscillator eigenmodes get excited when the gain is equal to or greater than the total loss, and the round-trip phase shift is an integer multiple of 2$\pi$, and high gain values would excite chaotic SWs in SWAROs~\cite{wu2009chaos, wuhagerstrom2011tuning}.
On the other hand, continuous RF injection signals were found to suppress the chaotic SW response~\cite{Grishin2008620, Grishin2013gen}.

Our study focuses on the response of a SWARO driven in the neighbourhood of a dipole gap, with continuous RF signals under varying gain conditions.
The spectra recorded by tuning the drive frequency at constant power and gain showed two types of sidebands. 
We also observed spectral suppression, sideband narrowing, and injection pulling when we drove the SWARO with relatively high-amplitude RF signals.
\section{Experimental results}
The transmission characteristics of the YIG film placed on a pair of microstrip line antennas are presented in Section~\ref{subsec:swtx}.
We then constructed a SWARO by connecting this device and a variable gain amplifier in a ring topology and applied a drive signal to the SWARO.  Section~\ref{subsec:nonlin_drivenosc} discusses the driven dynamics of the SWARO.

\subsection{MSSW dipole gaps in a yttrium iron garnet (YIG) film}
\label{subsec:swtx}
\begin{figure}
    \centering
    \includegraphics[width=0.7\linewidth]{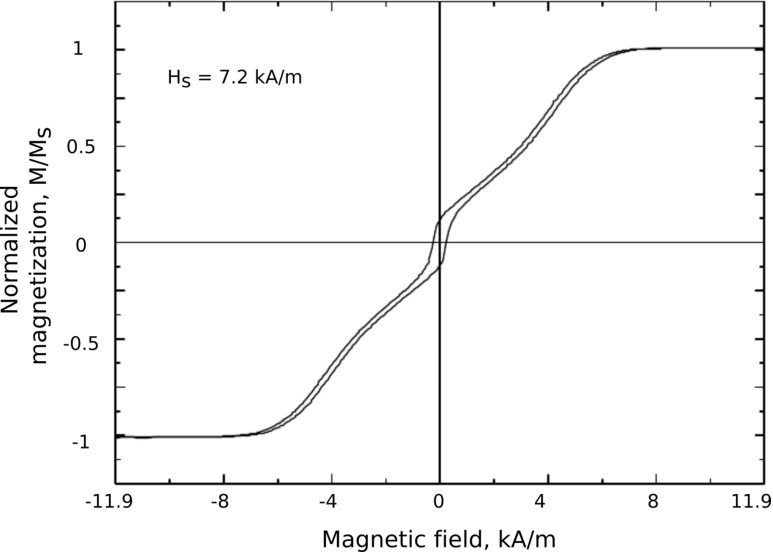}
    \caption{Hysteretic behaviour of in-plane magnetized YIG film sample. The sample gets saturated for any external field, $H_\text{ext} \ge 7.2\text{~kA/m}.$}
    \label{fig:yighyst}
\end{figure}
The YIG film has a size 18~mm $\times$ 14~mm $\times$ 6.9~$\mu$m, and a saturation magnetization, $M_\text{s}$ = 138.46~kA/m.
Any in-plane external magnetic field, \mbox{$H_\text{ext} \ge \text{7.2~kA/m}$}, is sufficient to saturate the film, according to the hysteresis curve given in Fig.~\ref{fig:yighyst}.
\begin{figure}[!h]
    \centering
    \includegraphics[width=0.55\linewidth]{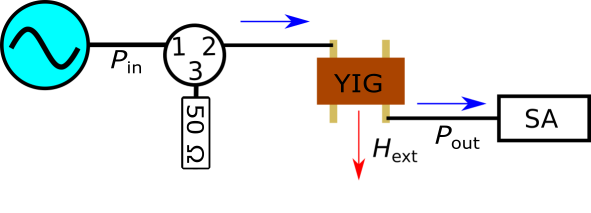}
    \caption{Electrical transmission spectroscopy setup to detect propagating MSSWs.}
    \label{fig:swtxsetup}
\end{figure}
Fig.~\ref{fig:swtxsetup} shows the electrical transmission spectroscopy setup that measures the MSSW transmission characteristics.
The film was kept on a pair of $\mu$-strip line antennas, separated by a distance of about 9~mm, with a magnetic field applied along the width of the YIG film.
\begin{figure}[!h]
    \centering
    \includegraphics[width=0.7\linewidth]{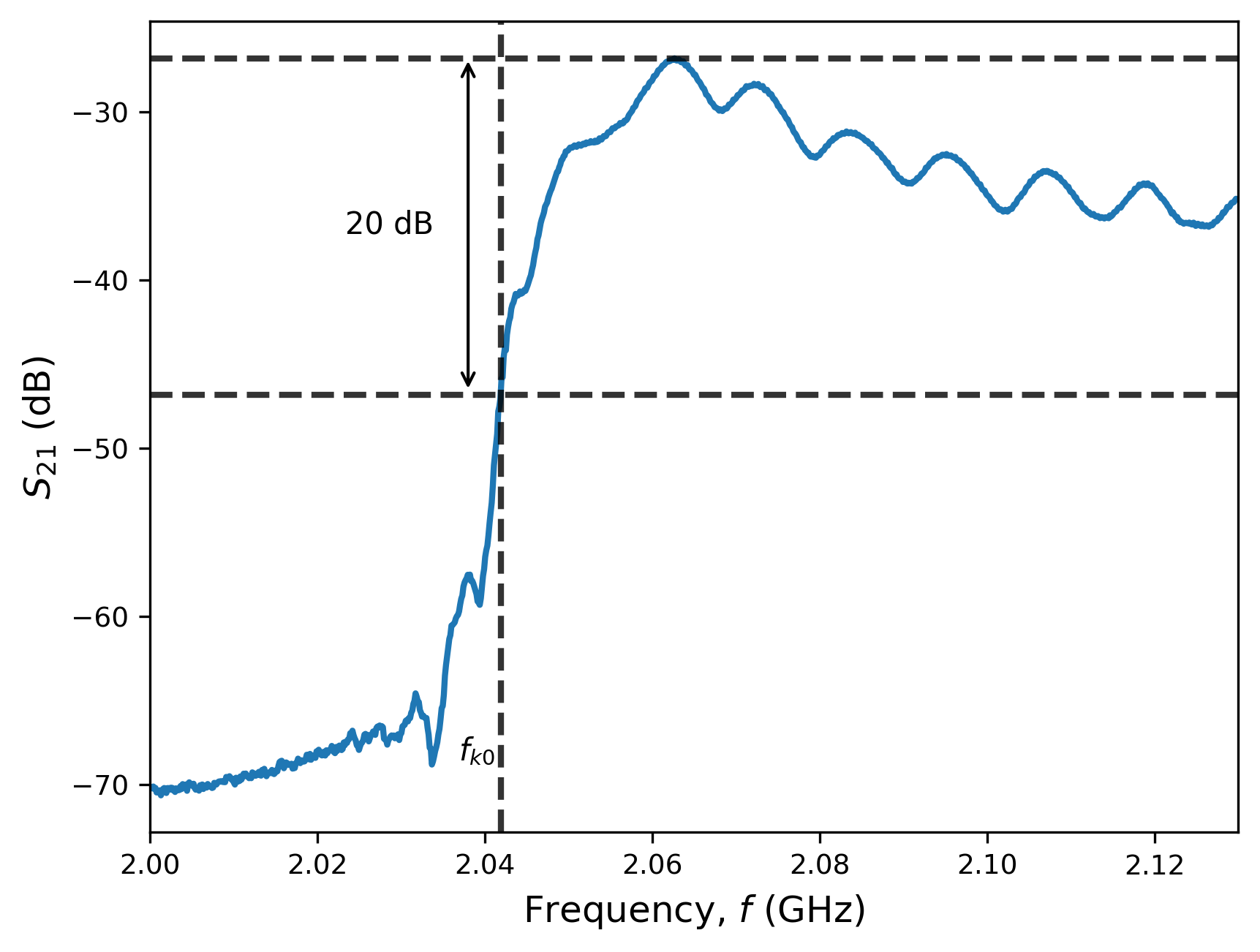}
    \caption{$S_{21}$ curve for propagating MSSWs. The lower edge of the MSSW manifold ($f_{k0}$) is marked with a vertical dashed line.}
    \label{fig:S21}
\end{figure}
A signal of strength -10~dBm was fed to the excitation antenna, and the signal frequency was swept from 2 to 2.13~GHz in steps of 130~kHz.
The transmission coefficient, $S_\text{21}$ shown in Fig.~\ref{fig:S21}, was obtained by a spectrum analyzer connected to the receiver antenna. 
We observe multiple dips in the passband, separated from each other by approximately 10~MHz. 
These dips are indicative of dipole gaps~\cite{Schilz1972}.
\begin{figure}[!h]
    \centering
    \includegraphics[width=0.7\linewidth]{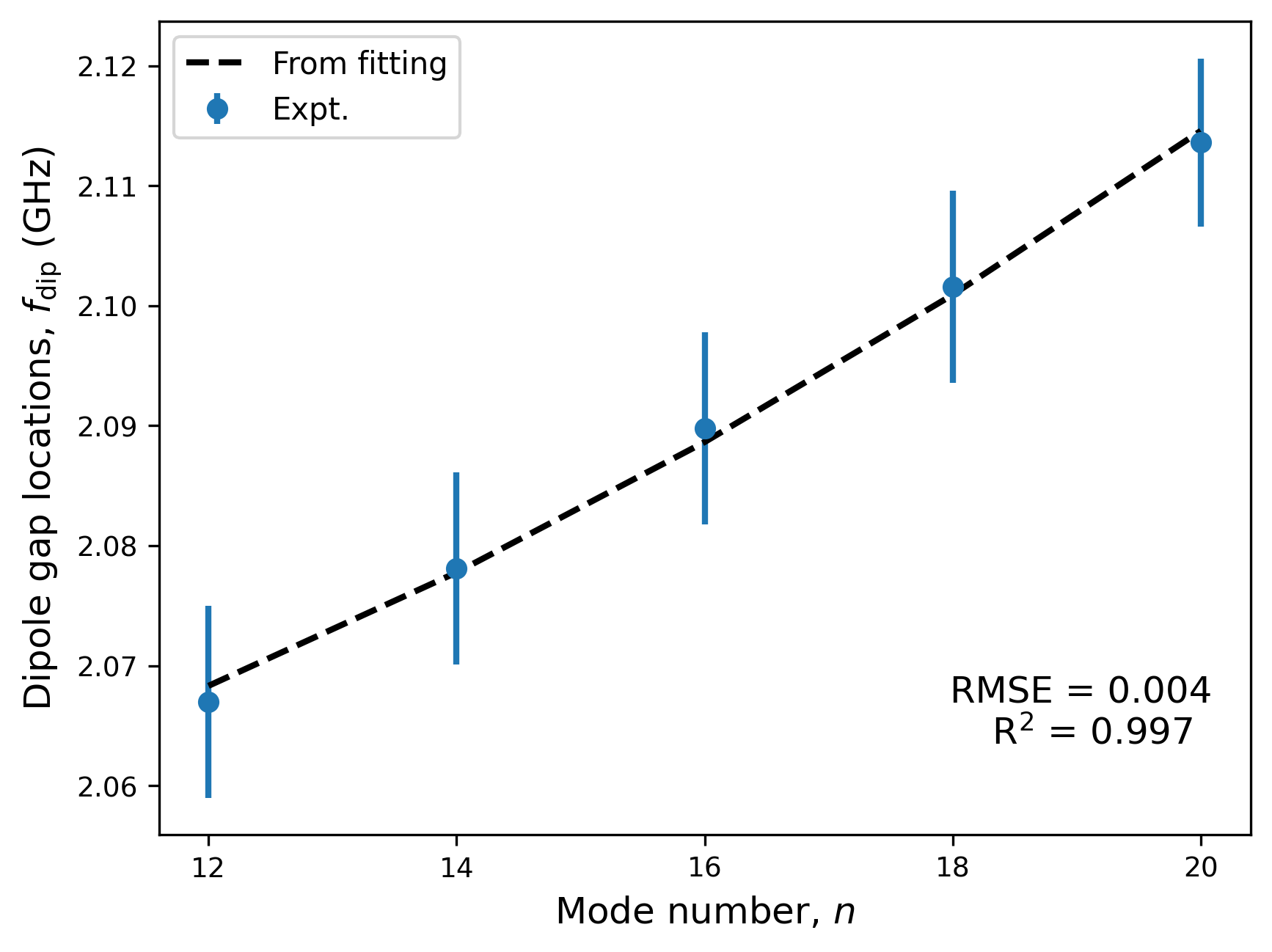}
    \caption{Experimentally obtained dipole gap frequencies at $H_\text{eff} \cong \text{21.09~kA/m}$. The dashed line is obtained from the least square fitting. The error bars indicate the uncertainty in the experimental measurement of dipole gap frequencies.}
    \label{fig:dipoloc_vs_n}
\end{figure}
We calculated the effective magnetic field, $H_\text{eff}$ from $S_{21}$ and then we used this field value along with the observed dipole gap frequencies to estimate the film's exchange stiffness, $A_\text{ex}$.
The lower edge of the MSSW manifold was assumed to be at $f_{k0} = \text{2.0419~GHz}$, 20~dB lower than the maximum value of $S_{21}$, as shown in Fig.~\ref{fig:S21}. 
In our experimental setup, the YIG film experiences  an effective magnetic field $H_\text{eff}$, estimated as
\begin{equation}
    \label{eq:k0freq}
    H_\text{eff} = M_\text{s} \left(-0.5 + \sqrt{0.25 + \left(\frac{f_{k0}}{f_\text{M}}\right)^2}\right) \simeq 21.09\,\text{kA/m},
\end{equation}
where, $f_\text{M} = |\gamma| \mu_\text{0} M_\text{s} / 2 \pi = \text{4.8738~GHz}$. 
$\gamma$ and $\mu_\text{0}$ are the gyromagnetic ratios for electron and free space magnetic permeability, respectively.
\begin{align}
        f_\text{dip}^2 &= f_\text{M}^2 \left(\Omega_\text{H} + K_n^2\right)\left( \Omega_\text{H} + K_n^2 + 1\right) \nonumber\\
        \label{eq:curve_fit}
        &= f_\text{M}^2 \left(\Omega_\text{H} + q^2 (n_0 + 2 m)^2\right) 
        \left(\Omega_\text{H} + q^2 (n_0 + 2 m)^2 + 1\right),
\end{align}
where, \mbox{$K_n = l_\text{ex}k_n = (l_\text{ex} \pi / t) n = qn = q(n_0 + 2 m)$}, $n_0$ is the mode number of the first dipole gap, \mbox{$l_\text{ex} = \sqrt{2 A_\text{ex} / \mu_\text{0} M_\text{s}^\text{2}}$} is the magnetostatic exchange length, and $t$ is the thickness of the film \cite{Schilz1972, Kalinikos1986}. 
\begin{table}[!htbp]
\centering
\caption{\label{tab:fit_param} Parameters estimated from fitting operation}
\begin{tabular}{cc}
\hline
   \textbf{Parameters} & \textbf{Values}\\
    \hline
    $n_0$ & 12\\
    $A_\text{ex}$ & 1.4~pJ/m\\
    $l_\text{ex}$ & 10.8~nm\\
    \hline
    \end{tabular}
\end{table}
We used a least-square algorithm to fit $m$ (independent variable) and average dipole gap frequencies (dependent variable) to Eq.~(\ref{eq:curve_fit}).
We allowed $m$ to take integer values ranging from 0 to 4 and treated $n_0, A_\text{ex}$ as fitting parameters.
The algorithm reached minimum error when $n_0 \approx 12$ and \mbox{$A_\text{ex} \approx$ 1.4~pJ/m}, as given in Table~\ref{tab:fit_param}.
\subsection{Nonlinear behaviour of spin wave active-ring oscillator}
\label{subsec:nonlin_drivenosc}
\begin{figure}[!h]
    \centering
    \includegraphics[width=0.5\linewidth]{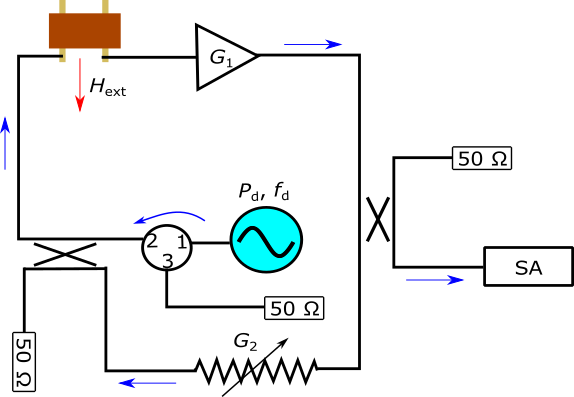}
    \caption{Schematic of SWARO with drive source. $P_\text{d}, f_\text{d}$ are drive power and frequency, respectively. $G_2$ is variable negative gain in the ring oscillator. SA stands for the spectrum analyzer.}
    \label{fig:ringoscsetup}
\end{figure}
The ring oscillator circuitry consists of a YIG film, a constant gain block amplifier ($G_1$), two directional couplers, a spectrum analyzer, a variable switch attenuator ($G_2$) and an external microwave source (driver oscillator), as shown in Fig.~\ref{fig:ringoscsetup}.
The two directional couplers inject the drive signal and couple out about 1\% of microwave power generated within the SWARO.
In our experimental setup, $G_1$ is a constant positive gain, and $G_2$ represents a tunable negative gain in the ring oscillator.
\begin{figure}
    \centering
    \includegraphics[width=\linewidth]{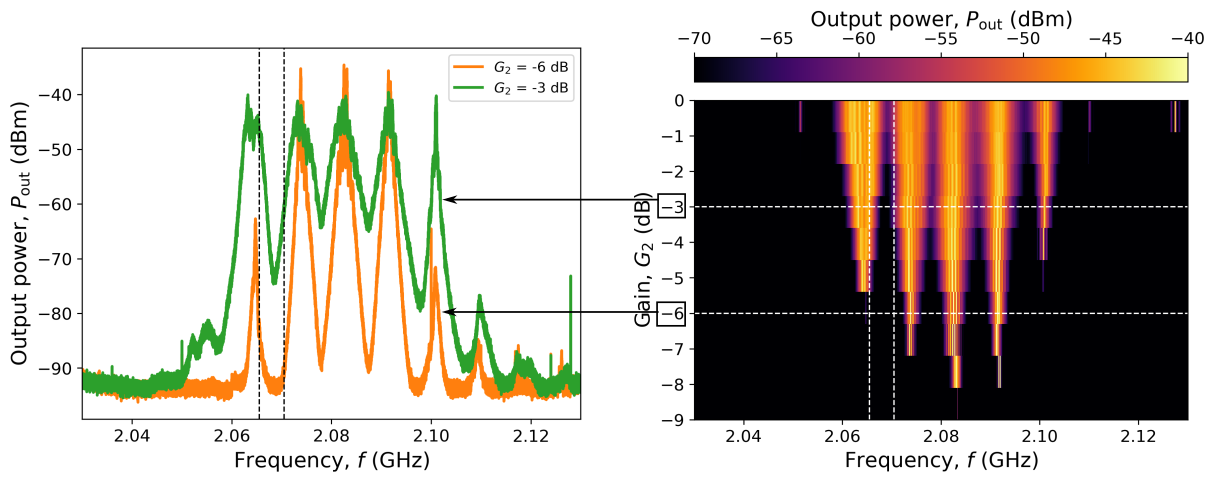}
    \caption{Spectral output from the free-running SWARO for different gain values. Two vertical dashed lines mark the observation window from 2.0655 to 2.0705~GHz.}
    \label{fig:sgspec}
\end{figure}
Using a spectrum analyzer, we measured the output spectrum from the ring oscillator with $G_2$ varied by step-size of 1~dB and no drive power, i.e. $P_\text{d} = 0\,\text{W}$ (free-running condition).
The SWARO mode with the lowest decay rate is excited when the total ring gain equates with the net loss. 
Other modes with comparatively higher decay rates are also generated with increasing $G_2$. 
The observed free-running spectra are plotted in ($G_2, f$) parameter space, in Fig.~\ref{fig:sgspec}, and we find that MSSWs near the dipole gaps are excited at higher $G_2$ values.
Hence, it is reasonable to claim that the dipole gaps sit between two neighbouring oscillator modes.
\begin{figure}[!h]
    \centering
    \includegraphics[width=\linewidth]{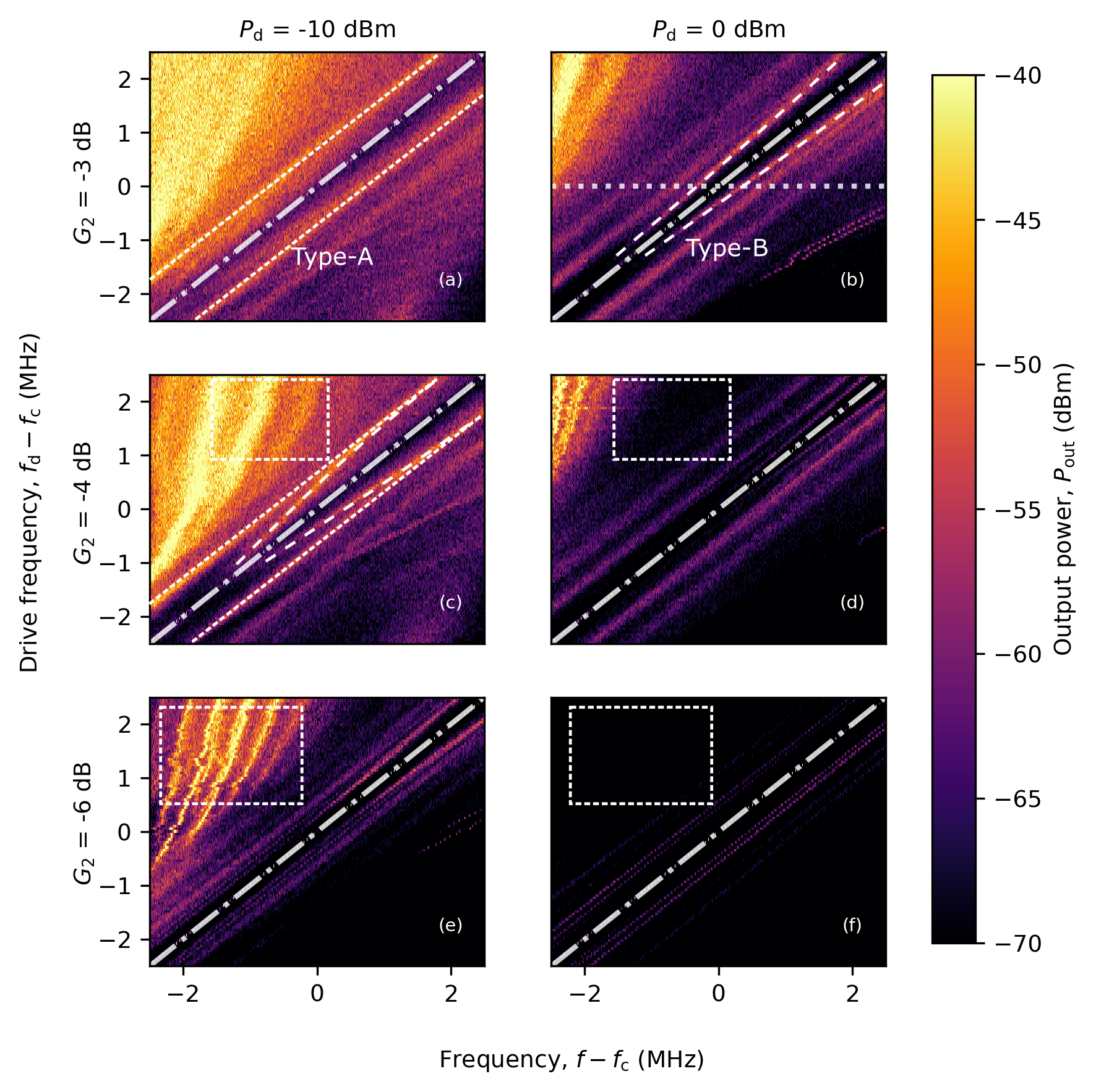}
    \caption{Spectral output of the SWARO driven at different $G_2$ and $P_\text{d}$. $G_2$ has three values, i.e. -6, -4 and -3 dB. $f_\text{c} = \text{2.068~GHz}$ is first dipole gap location. The dashed rectangles highlight the spectral suppression phenomenon. The diagonal dash-dotted lines are the drive frequency lines.}
    \label{fig:drivenspec}
\end{figure}
The first dipole gap at 2.068~GHz is labelled as $f_\text{c}$, refer to Fig.~\ref{fig:S21}. 
Spin waves do not propagate efficiently near a dipole gap, resulting in localized energy accumulation that, in turn, initiates various nonlinear effects~\cite{prabhakar1998auto, WU2010163}.
Our study is focused on a 5~MHz wide region around the first dipole gap $f_\text{c}$.
The drive frequency, $f_\text{d}$, was swept over this frequency range with a step-size of 50~kHz.
We recorded output spectra for different combinations of $f_\text{d}, P_\text{d}$ and $G_2$ values.
In Fig.~\ref{fig:drivenspec}, horizontal line scans on any surface plot are these spectra. 
As expected, nonlinear interactions dominate at higher $G_2$ values, and
we observe spectral suppression at high $P_\text{d}$ values~\cite{WU2010163}.
Fig.~\ref {fig:drivenspec}(c, e) show a yellowish-bright region above the diagonal drive frequency line. 
This region corresponds to the MSSW response greater than -45~dBm.
Fig.~\ref {fig:drivenspec}(d, f) show that this broad MSSW spectrum is suppressed by increasing $P_\text{d}$ by 10~dB.
The dashed rectangles in Fig.~\ref {fig:drivenspec}(c, d, e, f) highlight the large drive power induced spectral suppression phenomenon.
In addition, each surface plot in Fig.~\ref{fig:drivenspec} shows that weak sidebands accompany the drive signal. 
These sidebands can be classified into the following: 
\begin{enumerate}
    \item type-A: the distance between the sidebands and drive frequency remains constant.
    \item type-B: the sidebands move away from the diagonal drive frequency line as $f_\text{d}$ increases.
\end{enumerate}
Fig.~\ref{fig:drivenspec} shows that type-A sidebands are present for all the values of $G_2$, $f_\text{d}$ and $P_\text{d}$.
Type-B bands can be seen in Fig.~\ref{fig:drivenspec}(b, c).
\begin{figure}
    \centering
\includegraphics[width=0.7\linewidth]{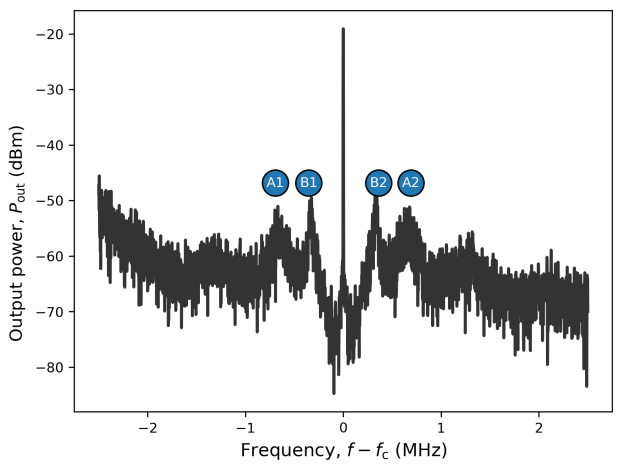}
    \caption{Power spectrum from the SWARO at $P_\text{d} = \text{0~dBm}$, $f_\text{d} = f_\text{c} = \text{2.068~GHz}$, and $G_2 = \text{-3~dB}$ (denoted by the horizontal white dashed line in Fig.~\ref{fig:drivenspec}(b)).}
    \label{fig:dswarospectrum_at2db0dbm}
\end{figure}
We extract a power spectrum from the spectral surface in Fig.~\ref{fig:drivenspec}(b) by fixing $f_\text{d} = f_\text{c} = \text{2.068~GHz}$ (denoted by the horizontal white dotted line) and plot it in Fig.~\ref{fig:dswarospectrum_at2db0dbm}.
Four sidebands can be seen, and they are marked with alphanumerals: A1, B1, B2 and A2. 
Peaks A1 and A2 are each approximately 0.6~MHz distant from the drive frequency.  
Peaks B1 and B2 are situated at $f_\text{d} - f_\text{c} \cong \pm\text{0.3~MHz}$ and shift towards peaks A1 and A2 with increasing drive frequency, respectively.
\begin{figure}[!h]
    \centering
    \includegraphics[width=0.7\linewidth]{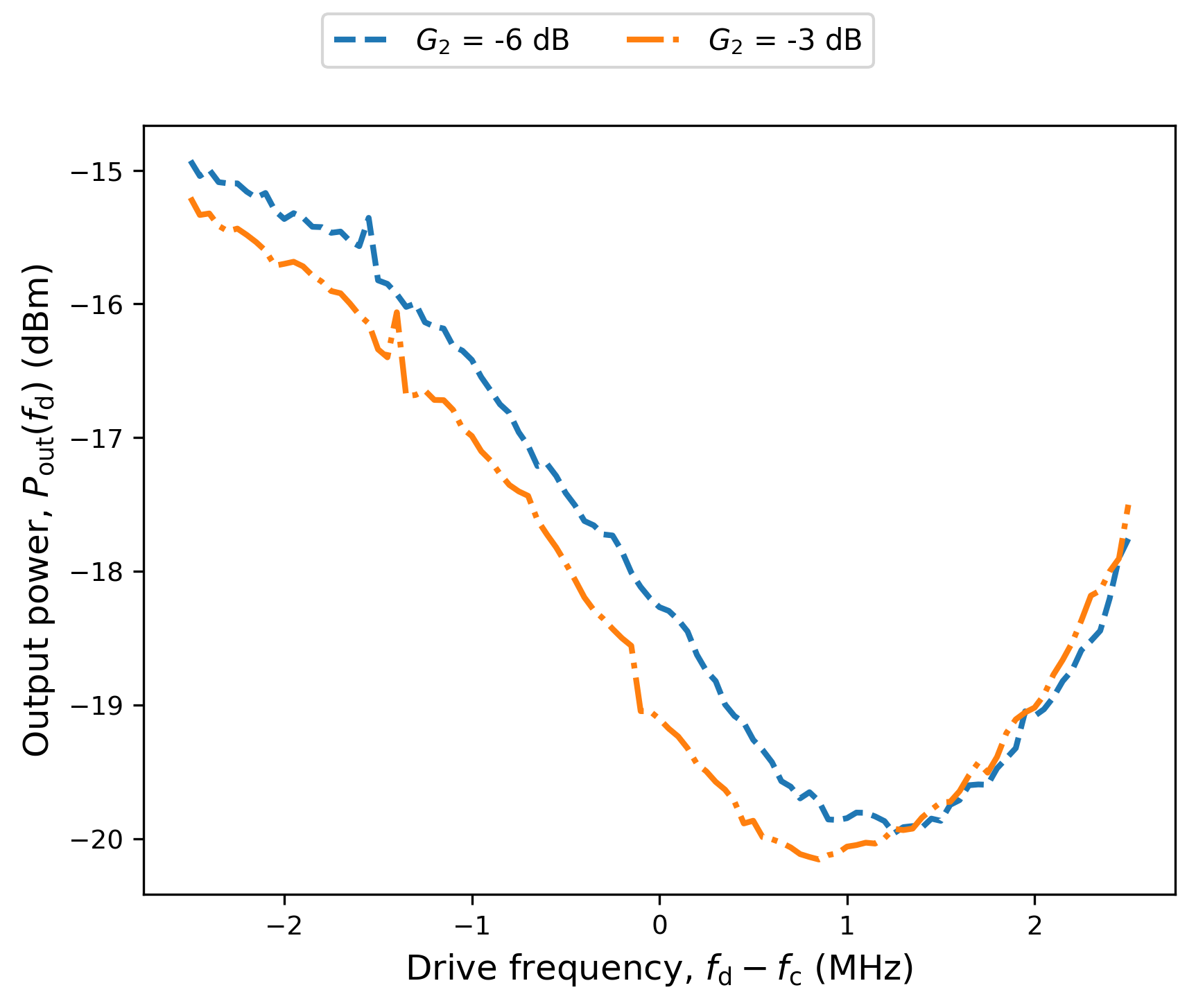}
    \caption{The power output from the SWARO at $f_\text{d}$ for two different values of $G_2 = \text{-6 and -3~dB}$ (i.e. along the dash-dotted lines in Fig.~\ref{fig:drivenspec}(b, f))are plotted as function of $f_\text{d}$. The $P_\text{d}$ was fixed at 0~dBm.}
    \label{fig:Sdatfd}
\end{figure}
We also extract the power along the diagonal dash-dotted line shown in Fig.~\ref{fig:drivenspec}(b, f) and plot it as a function of $f_\text{d}$ in Fig.~\ref{fig:Sdatfd}.
$P_\text{out}(f_\text{d})$ montonically decreases for \mbox{$f_\text{d} - f_\text{c} < \text{1~MHz}$} and begin to increase afterwards.
At $f_\text{d} - f_\text{c} \cong \text{1~MHz}$, we observe $\sim$5~dB reduction.
\begin{figure}[!h]
    \centering
    \includegraphics[width=0.8\linewidth]{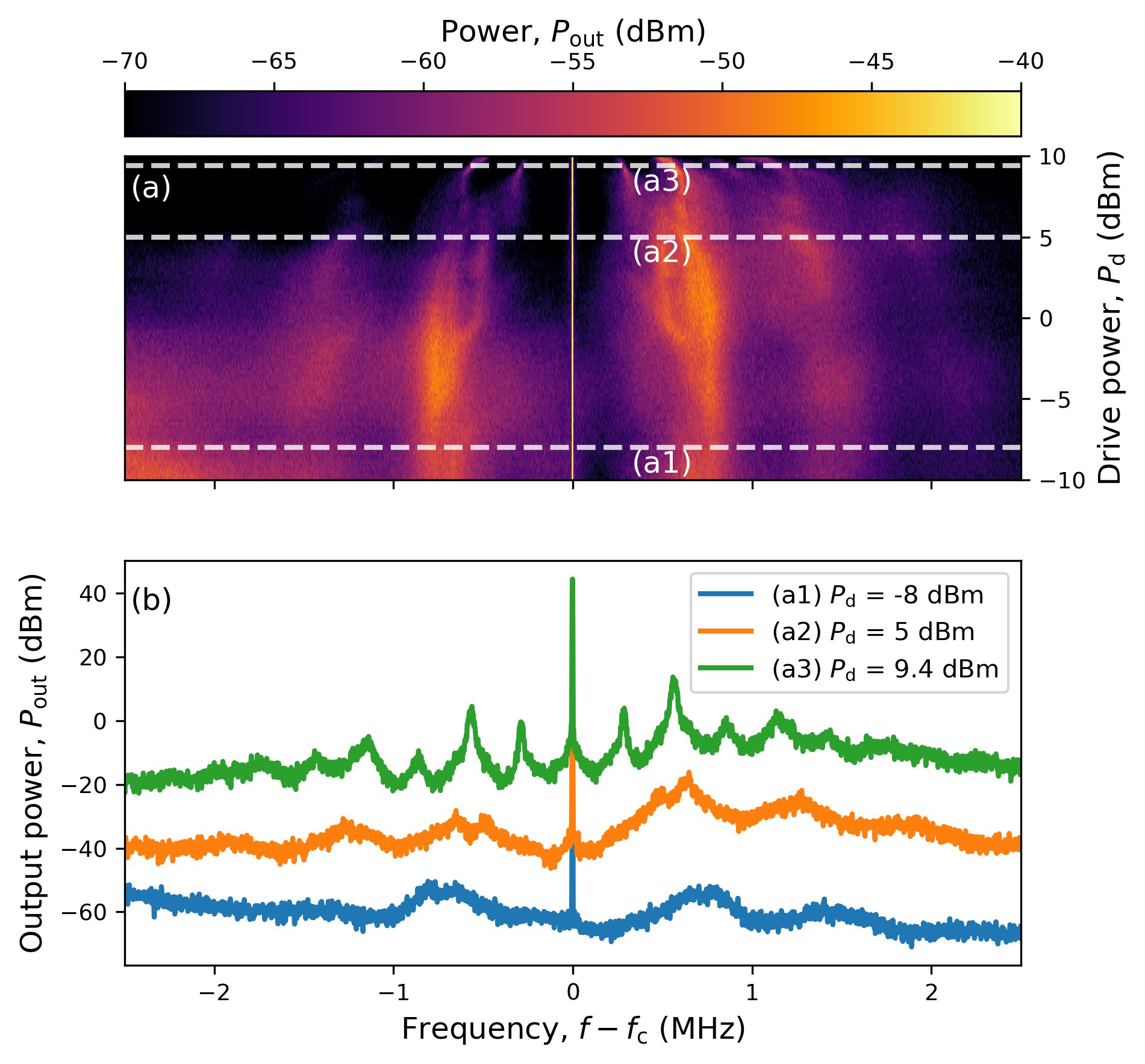}
    \caption{(a) Evolution of output spectrum from the driven oscillator with $G_2 = \text{-2~dB}$ at $f_\text{d} = \text{2.068~GHz}$. (b) Power spectra corresponding three different drive strengths, $P_\text{d} =$ (a1) -8~dBm, (a2) 5~dBm, and (a3) 9.4~dBm at same $G_2$ and $f_\text{d}$ values.}
    \label{fig:injlock_2dB}
\end{figure}

We now study the effect of increasing $P_\text{d}$ over the spectral response of driven SWARO at a fixed $f_\text{d}$ and $G_2$.
$P_\text{d}$ was varied from -10 to 10~dBm with a step-size of 0.2~dB at $f_\text{d} = f_\text{c} = \text{2.068~GHz}$ and $G_2 = \text{-2~dB}$.
Fig.~\ref{fig:injlock_2dB}(a) shows a surface map created by capturing spectra for each drive power.
We extract three power spectra corresponding to $P_\text{d} = \text{-8, 5, and 9.4~dBm}$, marked as dashed horizontal lines on the surface plot, refer to Fig.~\ref{fig:injlock_2dB}(b).
At $P_\text{d} = \text{-8~dBm}$, we observe two sidebands at $\Delta f = f - f_\text{c} \approx \pm\text{0.6~MHz}$.
As we increase $P_\text{d}$, the sideband at $f_\text{c} - \Delta f$ is attenuated while the power is redistributed at $f_\text{c} + \Delta f$.
Around $P_\text{d} = \text{7.5~dBm}$, multiple narrower sidebands appear around $f_\text{d}$.
These bands are pulled towards the $f_\text{d}$ as $P_\text{d}$ further increases.
\begin{figure}
    \centering
    \includegraphics[width=0.6\linewidth]{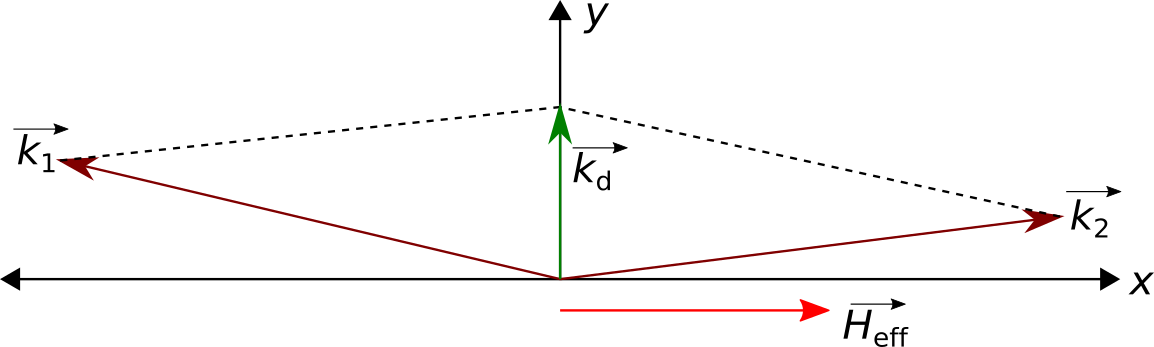}
    \caption{Vector diagram depicting momentum conservation for three-wave scattering process.}
    \label{fig:3magnon_vecdiag}
\end{figure}

The three-wave scattering processes (first-order parametric instability) were well studied in the case of the nonlinear MSSW dynamics~\cite{Cherkasski2010, Boardman198834}.
The energy-momentum conservation for such processes is as follows:
\begin{subequations}
    \begin{align}
     f_\text{d}(k_\text{d}) &= f_1(k_1) + f_2(k_2),\label{eq:3wave_fconsv}\\
     \myvec{k}_\text{d} &= \myvec{k}_1 + \myvec{k}_2,\label{eq:3wave_kconsv}
    \end{align}
\end{subequations}
where, ($f_\text{d}$, $\myvec{k}_\text{d}$) corresponds to MSSW mode. 
($f_1$, $\myvec{k}_1$) and ($f_2$, $\myvec{k}_2$) are parametrically excited BVSW mode pair.
Fig.~\ref{fig:3magnon_vecdiag} shows the momentum conservation in the three-wave interaction.
From Eq.~(\ref{eq:3wave_fconsv}), we can state that the three-wave interaction occurs if, 
\begin{equation}
    f_{k0} \ge f_1|_\text{min} + f_2|_\text{min} = 2 f_\text{H},\,\,\text{or}\,\, f_{k0} \le 2 f_\text{H} \le f_{k\infty}\label{eq:3wave_freq_lim}
\end{equation}
where, $f_{k\infty} = f_\text{M} \sqrt{\Omega_\text{H} (\Omega_\text{H} + 1) + 0.25}$ is the upper bound of MSSW manifold and $f_\text{H} = f_\text{M} \Omega_\text{H}$ is the lowest BVSW frequency~\cite{Jun1997}. 
The inequality relation in Eq.~(\ref{eq:3wave_freq_lim}) also provides a range of $H_\text{eff}$ where three-wave interaction is allowed, refer to Table~\ref{tab:3wave_heff_lim}.
\begin{table}[!h]
        \centering
        \caption{Effective field range for three-wave interaction.}
        \label{tab:3wave_heff_lim}
        \begin{tabular}{ccc}
        \hline
         \textbf{$H_\text{eff}$-range} & \textbf{$f$-range} &  \textbf{Three-wave scattering}\\
         \hline
           $H_\text{eff} \le M_\text{s}/3$  & $[f_{k0}, f_{k\infty}]$ & Allowed \\
           $M_\text{s}/3 < H_\text{eff} \le M_\text{s}/2$ & $[f_{k0}, 2f_\text{H})$ & Not allowed \\
           $M_\text{s}/3 < H_\text{eff} \le M_\text{s}/2$ & $[2f_\text{H}, f_{k\infty}]$ & Allowed \\
           $H_\text{eff} > M_\text{s}/2$ & - & Not allowed\\
           \hline
        \end{tabular}
    \end{table}
In our experiments, \mbox{$H_\text{eff} < M_\text{s} / 3 = \text{46.15~kA/m}$} and the experimental frequency range is above $2 f_\text{H} = \text{1.48~GHz}$.
Hence, we believe that the weak sidebands appear owing to three-wave interaction.
\section{Summary}
We have explored the nonlinearity in an MSSW configuration with a YIG film in a microwave ring oscillator near a dipole gap.
We injected microwave signals in the dipole gap to observe a rich dynamical landscape of nonlinear MSSW excitations as we changed the gain of the ring oscillator. 
We observed that the power in the injected signal, $P_\text{d}$, controls the extent of the nonlinear interactions. 
At relatively low drive power, sidebands get excited. 
We believe that the three-wave scattering plays a role in side-band generation.
As the power increases, the sidebands are pulled towards the drive frequency.
In conclusion, our study provides insight into the complex interplay between drive signal, oscillator gain, and spin-wave nonlinearity in SWAROs.
\section*{Acknowledgments}
We thank N. Bilanuik and D. D. Stancil for the microstrips, the Department of Science and Technology for support vide sanction order IWT/UKR/P-28/2018 and acknowledge the bilateral Indian–Ukrainian S\&T cooperation project (M/11-2020).
\bibliographystyle{elsarticle-num-names} 
\bibliography{REF.bib}

\end{document}